\documentclass[aps,amsmath,amssymb,twocolumn,showkeys,preprintnumbers]{revtex4}
\usepackage[dvips]{graphicx,color}
\usepackage{times}
\usepackage{mathrsfs}
\usepackage{natbib}
\usepackage{amsmath}
\usepackage{graphicx}
\usepackage{epsfig}
\usepackage{dcolumn}
\usepackage{bm}
\definecolor{r}{rgb}{1.0,0,0}

\begin{document}
\title{Searching for PeV neutrinos from photomeson interactions in magnetars}
\author{Rajat K. Dey}
\email{rkdey2007phy@rediffmail.com}
\author{Sabyasachi Ray}
\email{methesabyasachi@gmail.com}
\author{Sandip Dam}
\email{sandip_dam@rediffmail.com}
\affiliation{Department of Physics, University of North Bengal, Siliguri, West Bengal, INDIA 734 013}

\begin{abstract}

In this paper we estimate the flux of PeV neutrinos and gamma-rays from magnetar polar caps, assuming that ions/protons are injected, and accelerated in these regions and interact with the radiative background. The present study takes into account the effect of the photon splitting mechanisms that should modify the radiative background, and enhance the neutrino and gamma-ray fluxes at PeV energies, with a view to explain the PeV neutrino events detected in IceCube. The results indicate that in near future, possibility of any significant excess of neutrino events from a magnetar in Milky Way is extremely low. Further, we suggest that the simultaneous observation of neutrinos and gamma-rays at Earth from expanded {\lq Gen2\rq} IceCube detector and/or High Altitude Water Cherenkov Observatory would provide opportunities to explore the possible origin of very high energy neutrinos and gamma-rays. 
  
\end{abstract}

\maketitle

\section{Introduction}

The detection of very high energy (VHE) cosmic neutrinos with the IceCube detector has recently opened up a whole new window on the energetic Universe. Although the significance of the IceCube excess over the atmospheric component is quite high, detection of point sources (and thus individual source classes) has of yet eluded the neutrino astrophysics community. In this context, the present paper in principle proposes and explores a viable physical model of the interactions taking place in source regions for the prediction of astrophysical neutrinos and accompanying gamma-rays. 

Probable candidates of VHE neutrinos and gamma-rays include various extragalactic sources, viz active galactic nuclei (AGNs), gamma-ray bursts (GRBs), or star-burst galaxies. The VHE neutrinos from these sources can travel over extragalactic distances retaining their directional information unaffected and not experiencing strong absorptions. However, the extragalactic PeV gamma-ray component will be attenuated significantly due to pair-production processes in cosmic background radiations and no longer remain as important cosmic messengers in order to explore their origin. A continuous effort is being made by the 1 km$^3$ IceCube neutrino observatory in order to detect such astrophysical neutrinos \citep{hal14}. The IceCube detector has recently observed two VHE neutrinos above $1$ PeV satisfying all selection criteria in two years data collection, might be of astrophysical origin \citep{aar13}.

The possible origin of these PeV neutrino events have been discussed by many in recent times \citep{lah13, aar14} but still remains a topic of much speculation. A fundamental question is therefore being raised: what classes of astrophysical objects could accelerate hadrons to very high energies, and in which types of interactions are neutrinos then produced?

In this situation, acceleration of protons in the vicinity of the surface of relatively young local neutron stars with super strong magnetic fields ($\rm{B} \sim 10^{14-16}$ G), widely known as magnetars, is in fact supposed to be a possibility. Although, it may be not a highly probable process and as of yet not observationally supported. The subsequent production of photomesons by interactions with radiative background proceeds, and that has already been studied by many authors. In this paper, we have proposed an additional contribution to the target photon fields for photomeson production from the photon splitting mechanism. The photon splitting process is expected to modify the radiative background, and enhance the neutrino and gamma-ray fluxes at PeV energies from the object. The two astrophysical PeV neutrino events detected in IceCube could have their origin within a young local magnetar environment. It is also found that our model dependent calculation on the PeV neutrino flux limits the upper bound of the accompanying gamma-ray flux at these energies. 

Recently identified magnetar candidates are traditionally known to be either anomalous X-ray pulsar (AXPs) or soft gamma-ray repeaters (SGRs), mainly considered as high-energy X-ray emitters \citep{ola14, mer13}. Like pulsars \citep{lin05, lin06, bha09}, it has been learned that if magnetars polar caps with $\bf{\Omega \cdot \rm{\bf{B}}} < 0$ can accelerate protons/ions, TeV neutrinos and gamma-rays could be produced via photomeson productions due to interactions of these accelerated protons with UV/X-radiations close to polar caps regardless of their relatively slow rotations \citep{zha03}. 

The phase diagram of a magnetar describes the variation of star's magnetic field ($\rm B$) with spin period ($\rm P$). The diagram $\rm{P} - \rm{B}$,  corresponds a phase in which the star could pass a neutrino-loud regime for some suitable set of parameters characterizing its evolution \citep{zha03}. This present work shows that if the spin-down power of a local magnetar for a favorable set of $\rm P$ and $\rm B$ at a particular evolutionary phase is consumed to accelerate protons, then the object might emit PeV muon neutrinos ($\nu_{\mu}$) and gamma-rays through photomeson interactions between these protons in polar caps and star's radiative background. The radiative background of the star is believed to be filled mainly with soft ultraviolet (UV)-A or B photons that are in turn produced from the effect of photon splitting mechanism on magnetar's unmodified radiative background (i.e. UV-C/X radiation). It is worth noting that this type of magnetars have not been reported from observation yet; they are predicted theoretically only \citep{zha03}.
   
The process, $\gamma \rightarrow e^{+} + e^-$ has so far been applied in polar caps with strong magnetic environment to account for the energy loss by a high-energy $\gamma$-ray photon. Another process, $\gamma \rightarrow \gamma + \gamma$, called photon splitting is expected to take place in super strong magnetic-field regions of polar caps \citep{har97, chi12}. The effect of the photon splitting mechanism on star's own radiation field prior to photomeson interactions is a principal step that could be applied to available astrophysical settings in magnetars. 

This paper is organized as follows. In Sect. II, the possible implications of the photon splitting process to the modification of the radiative background is discussed. In Sect. III,  we provide an overview of the physical model following its detailed description. In the same section, we estimate the expected flux of PeV muon neutrinos and gamma-rays from a local magnetar candidate based on this model. A brief discussion of the prospects for detection of such high-energy neutrinos and gamma-rays by the ongoing and the future experiments is given in Sect. IV. In Sect. V, we summarize our conclusions. 

\section{Photon splitting process in the magnetosphere}

The photon splitting, a QED process that splits a high-energy photon into pair of low energy photons in presence of a pure magnetic field and/or magnetized plasma. The effect leaves some important signatures in astrophysical environments e.g. magnetars, gamma-ray bursts etc. where magnetic fields approaching or even exceeding the quantum critical value, $\rm{B_{cr}} = 4.41 \times 10^{13}$ G. Indication of spectral cutoffs in the observed gamma-ray spectra of SGRs, and the radio quiescence of SGRs and AXPS could be explained by exotic processes (photon splitting cascades, merging etc.) under consideration \citep{bar95, tho95}

Generally, in a strongly magnetized plasma or vacuum, two more competing processes; photon merging and Compton scattering may onset along with the photon splitting \citep{har97, chi12}. No detailed calculation on the probabilities of these processes in the current scenario ($\rm{B} > \rm{B_{cr}}$ with plasma) is available except a numerical calculation used in \citep{chi12}. The photon splitting rate could be affected considerably by the magnetized plasma. The environment affects the rate by changing photo dispersion properties of the region. The ambient radiation field in the vicinity of a magnetar can be approximated to an environment equivalent to a magnetized rarefied plasma (inclusion of $\gamma - \gamma$ pairs in combination with $\rm{e}^{+} - \rm{e}^{-}$ pairs in super-strong field). On the other hand, a region of the magnetar's magnetosphere where high-energy gamma-rays and neutrinos are generated, is considered to be a strong magnetized vacuum i.e. environment without plasma. 

Photon splitting may occur via various possible channels which are determined by the electric field vector ($\bf{\rm{X}}$), momentum vector ($\bf{\rm{q}}$) associated with a photon, and the magnetic-field vector ($\bf{\rm{B}}$). The state \lq 1\rq  refers to a configuration where $\bf{\rm{X}}$ is perpendicular to the plane containing $\bf{\rm{q}}$ and $\bf{\rm{B}}$ while \lq 2\rq  corresponds the $\bf{\rm{X}}$ being parallel to their plane. For low energy background target UV-C photons ($\rm{T_{kin.}} \cong 0.1 - 0.2$ keV $<< \rm{m_{e}c}^{2}$) the only physical mode, $\gamma_{1} \rightarrow \gamma_{2} + \gamma_{2}$ is responsible for splitting and, thereby soften background photon spectra in the magnetized plasma \citep{chi12}. In addition, the photon merging channel, $\gamma_{2} + \gamma_{2} \rightarrow \gamma_{1}$ also plays an important role in the attenuation and softening of target photons. But in the absence of plasma, the physical mode, $\gamma_{1} \rightarrow \gamma_{1} + \gamma_{2}$ is the significant one for photon splitting over the channel, $\gamma_{1} \rightarrow \gamma_{2} + \gamma_{2}$. It should be however mentioned that the influence of the Compton scattering has negligible effect on softening of background target photons against photon splitting in magnetar model.

In magnetar model, conversion of $\pi^{0}$ into $\gamma$-rays generally occurs near the magnetic vacuum region of the magnetosphere 
where the magnetic-field intensity drops to $\rm{B_{cr}}$ or lower. As a consequence the influence of the main exotic process $\gamma_{1} \rightarrow \gamma_{1} + \gamma_{2}$ on the modification of the emergent PeV gamma-ray spectra in such a pure magnetic field essentially becomes unimportant.

\section{Emission of PeV neutrinos and gamma-rays}
 
\subsection{Photomeson production}

There has been a consensus among researchers in particle astrophysics over the last few decades that the VHE protons and/or heavier ions are injected, and accelerated in surrounding regions of cosmic accelerators (e.g. pulsars, nebulae, supernovae remnants, young magnetars). These accelerated ions then interact with the radiative background and/or the ambient matter. Subsequently VHE neutrinos and gamma-rays are generated through several reaction channels following dominant photomeson interactions in the radiative background as

\[{\rm p} + \gamma \rightarrow \Delta^{+} \rightarrow \left\{\begin{array}{ll}
{\rm p} + \pi^{o} \rightarrow {\rm p} + 2\gamma \\
{\rm n}\pi^{+} \rightarrow {\rm n} + {\rm e}^{+} + \nu_{\rm e} + \nu_{\mu} + \bar{\nu_{\mu}}
\end{array}
\right.  (A)\] 
										   
The final products of all neutrino flavours with gamma-ray keep the ratio approximately as ${\nu}_{\rm{e}} : {\nu}_{\mu} : {\nu}_{\tau} : \gamma = 1:2:0:2$ at the sources but the process of neutrino oscillation turns this into a ratio of ${\nu}_{\rm{e}} : {\nu}_{\mu} : {\nu}_{\tau} : \gamma = 1:1:1:2$ while observing at earth.

In magnetar's early life, the spin-down power is consumed to accelerate protons/ions, and the magnetic-field driven power supplies ambient photon targets. The kinematic threshold for photomeson interaction process in (A) is determined by the accessible photon energies in the radiative field. Most of the radiative target photons are of UV type, as characteristic of many stars (pulsars, nebulae, accreting objects), the kinetic energy of protons has to range from a few tens to hundreds of PeV.

For common magnetars like SGRs and AXPs, the surface magnetic field might be of the order of $10^{15}$ G. The photon splitting process is assumed to take place much above the polar caps (away from $\gamma$-ray generation point) where the star's magnetic field does not sharply reduce to $\sim 10^{12}$ G. Below the $\rm{B_{cr}}$, the magnetic field inhibits the photon splitting mechanism. Emission of VHE neutrinos and gamma-rays from these slowly rotating objects is described by the outer gap model \citep{che86}. In this gap closure scenario, these class of magnetars experience dominant gamma-ray energy degradation via the $\gamma - \rm{B}$ and the $\gamma - \gamma$ cascade formation, and the photon splitting process (although not included). Degraded gamma-ray energies fall below the cutoffs for direct observations viz. EGRET and Fermi-LAT \citep{bar01, ton10}. 

It is clear that the principal effect of the exotic photon splitting process is responsible to lower photon energies estimated at points away from a source with extreme physical conditions \citep{chi12}. But in the present work, it has been suggested and believed that the exotic process in a magnetar may also degrade energies of background target photons in the star's radiative field along with conventional pair-production loss i.e., $\gamma \rightarrow \rm{e}^{+} + \rm{e}^{-}$. Though in \citep{chi12}, the photon splitting process was studied around $\epsilon_{\gamma} \sim 0.51$ MeV, but its effect in pure magnetic field might be significant even at lower energies ($\epsilon_{\gamma} << \rm{m_{e}c}^{2}$). If so, a main fraction of the UV photons ($0.2 - 0.4$ keV) of the radiative background might convert into softer radiation comprising chiefly by UV-A and UV-B types with average photon energy $\approx 0.01$ keV. The accelerated protons in polar caps will interact with these modified UV photon targets via photomeson processes in open field line regions of the magnetosphere. Consequent upon, the PeV neutrinos and gamma-rays are generated and be detected on earth. 
       
\subsection{Physical model and photomeson threshold}

For young pulsars/magnetars with large spin-down power, acceleration of protons/ions has been considered by the widely used polar cap \citep{rud75, aro79, har98} and outer gap models \citep{che86} inside the magnetosphere. More clearly, charged particles undergo acceleration in the open field line zone near the magnetar's pole in the polar cap model. In the later model, acceleration of particles takes place in the empty gaps within the bounded region by the neutral and last open lines in the magnetosphere.

The neutron star (NS) remnant or merger that appears from massive binary NSs ($\rm{M} \sim 2\rm{M}_{\odot}$) coalescence, may form a millisecond magnetar with thin ejecta walls across polar caps in its very early phase. As the star receives huge angular momentum from the binary it possesses a rapid rotation at the moment of its birth. These magnetars also have super-strong magnetic fields \citep{zra13, dan92}. However, formation of a prompt black hole from the merging can't provide such a spin-down driven energy injection. On the other hand, detection of quiescent emission and flares from some magnetars (normally as slow rotators with $\rm{P} \sim 5 - 12$s) can only be interpreted in terms of magnetic-field driven power \citep{dan92, tho95}. Here, we suggest that in the evolutionary phase of a magnetar after NSs merger, the star may transit a state when spin-down power is comparable with its magnetic power, and spin period falls in the range $200 - 500$ ms \citep{zha03}. At this phase, the magnetar may induce photomeson processes in the polar cap that in turn produce PeV neutrinos and gamma-rays.     

Generally, the spin down power induces strong electric fields in polar caps where charged particles are injected, and experience acceleration, and directed towards the open field lines near the star's pole. The ultimate limit of potential drop of a magnetar with angular velocity $\Omega = 2\pi/\rm{P}$ corresponding to the induced electric field across the magnetic-field lines from the magnetic pole to the last line, that extends to infinity, is $\Delta \phi = \rm{B_{S}R}^{3}\Omega^{2}/2\rm{c}^{2}$ \citep{gol69}. In the expression, $\rm R$ denotes the radius of the magnetar, $\rm{B_{S}}$ be the strength of magnetic field at star's surface, and $\rm c$ is the speed of light. In the present work, the evolutionary phase of a magnetar is considered to be nearly analogous to a young pulsar except its magnetic-field intensity ($\rm{B_{S}} \equiv \rm{B}_{15} \times 10^{15}$ G) and the existence of $\gamma - \gamma$ pair cascades in addition to  $\rm{e}^{+} - \rm{e}^{-}$ pairs. The magnitude of the potential drop would be huge with a value of $7 \times 10^{21} \rm{B}_{15} \rm{P_{ms}}^{-2}$ volts. In magnetar's  polar caps, presence of $\rm{e}^{+} - \rm{e}^{-}$ pair cascades in the strong magnetic field may topple the electric field slightly along the field lines due to screening effect in comparison with pulsars. The presence of such $\rm{e}^{+} - \rm{e}^{-}$ pair cascades were validated by the results obtained from the observation of the crab and other pulsar wind nebulae (PWN). But the current understanding related to various aspects of the cascade formation is still incomplete and hence it needs more research.            

An extension of previous calculations for young pulsars \cite{lin05} and \cite{lin06} to magnetars reveals that protons or heavier ions undergo acceleration in the magnetar's polar caps attaining energies close to $10^{16} - 10^{17}$ eV, provided the magnetar's magnetic moment vector $\bf{\mu}$ and $\bf{\Omega}$ parameter satisfy the strong condition; $\bf{\mu}.\bf{\Omega} <0$. These VHE protons will interact with soft UV-A and UV-B photons close to the magnetar's polar caps, the $\Delta$ resonance state may form satisfying the kinematic threshold condition for the process in (A). The photomeson production threshold for a proton to reach the $\Delta^{+}$ state is something where the kinetic energies of the proton ($\epsilon_{\rm p}$) and UV-A/B photon ($\epsilon_{\gamma}$) would satisfy 

\begin{equation}
\epsilon_{\rm{p}}\epsilon_{\gamma} (1-\rm{cos}\theta_{\rm{p}\gamma}) \ge 0.3 \; \rm{GeV}^{2}, 
\end{equation}
where $\theta_{\rm{p}\gamma}$ is the incident angle between the proton and photon as measured in the laboratory frame. In a young magnetar, the typical photon energies near the polar caps are much smaller than a young pulsar because of intense magnetic field. The energy of the modified target photons is $2.8\rm{kT}_{\infty}(1+\rm{z_{g}}) \sim 0.01$ keV, where $\rm{z_{g}}$ $\sim 0.4$ being the gravitational red shift. Thus the proton threshold energy $\epsilon_{\rm{p},\rm{Th}}$ for the $\Delta^{+}$ resonance state ranges $\geq 3\times10^{16}$ eV.    

\subsection{The ion/proton flux interact in radiative fields}

Since one may expect that the approximate ion injection rate around the equatorial sector is the Goldreich-Julian rate \citep{gal94} and hence for a quasi-static magnetospheric environment the charge density near magnetar surface can be approximated to $\rho_{\rm{q}} \simeq \rm{eZn_{o}}$ when the Goldreich-Jullian density of ions at a radial distance $\rm{r}$ is equated to $\rm{n_{o}(r)}\equiv \rm{B_{s}R}^{3} \Omega /(4 \pi \rm{Ze c r}^{3})$ \citep{gol69}. There must be charge depleted gaps above stellar surface to induce acceleration where the density of ions may be parametrized as $\rm{f_{d}(1-f_{d}) n_{o}}$, with $\rm{f_{d}}$ ($ < 1$ for the modest depletion) is the depletion factor depending on adopted models and is unreliable one. The VHE proton flux emitted from the polar cap region would therefore be
              
\begin{equation}
\Phi_{\rm{PC}}\simeq \rm{c f_{d}(1-f_{d}) n_{o} A_{PC}}, 
\end{equation}

where $\rm{A_{PC}}$ denotes polar cap area, and it is $\eta_{\rm{A}} (4 \pi \rm{R}^{2})$ with $\eta_{\rm{A}}$ accounts the ratio of polar cap area to the magnetar surface area. Earlier calculations in \citep{lin05, lin06} for estimating proton/ion flux in pulsar's polar caps took the parameter $\eta_{\rm{A}}$ as unity. The characteristic polar cap radius can be given by, $\rm{r_{PC}}=\rm{R} (\Omega \rm{R/c)}^{1/2}$, and hence $\eta_{\rm{A}}$ takes the form $\Omega \rm{R/(4c)}$ \citep{bes93}. 

\subsection{The neutrino and gamma-ray fluxes on earth}

It is seen from the process in (A) that the charge-changing reaction goes on just $\frac{1}{3}$-rd of the reaction time, about three high-energy neutrinos (or a pair of $\nu_{\mu},\bar{\nu_{\mu}}$) will accompany with four high-energy gamma-rays on the average when a significant number of such reactions proceed successfully. The muon neutrinos and gamma-rays that are produced from charged and neutral pions via $\Delta$ resonance will be moving almost in nearly the original direction of protons. These neutrinos will arrive at earth without suffering any change in flux and energy. But, the high-energy gamma-ray flux might suffer a change due to QED effects in presence of strong magnetic field in magnetar's magnetosphere. Suppose, the factor $\rm{f_{s}}$ accounts such a modification in gamma-ray flux.

The accelerated ions in polar caps will suffer interaction in the UV-A/B dominated radiative background of a magnetar. For a young magnetar with surface temperature $\rm{T}_{\infty}$, the UV-C photon density in the vicinity to the star surface area is $\rm{n}_{\gamma}(\rm{R})=(\rm{a_{SB}}/2.8\rm{k})\left[(1+\rm{z_{g}})\rm{T}_{\infty}\right]^{3} $, $\rm{a}_{\rm{SB}}$ being the Stefan-Boltzmann constant. A numerical value of $\rm{n}_{\gamma}(\rm R)$ could be approximated as $9 \times 10^{19} \rm{T}^{3}_{0.1 \; \rm{keV}}$ (close to $\rm R$, $\rm{T}_{0.1 \; \rm{keV}} \sim 0.5$). The photon density increases due to splitting and at the same time will decrease with the increase of radial distance from the stellar surface. We assume that these two variations will keep the overall photon density nearly constant not much distance away from the star's surface.  The UV-C type photon spectra will reduce into soft UV-A/B spectra only. Now, the conversion probability for $\rm{p} \rightarrow \Delta^{+}$ via UV-A/B interaction along the distance from $\rm R$ to $\rm r$ is given by (LB) $\rm{P_{c}(r)}=1 - \rm{P(r)}$, where $\rm{dP/P(r)} =-\rm{n}_{\gamma}(\rm r) \sigma_{\rm{p}\gamma} \rm{dr} \simeq - \rm{n}_{\gamma}(\rm{R}) \sigma_{\rm{p}\gamma} \rm{dr}$ with $\epsilon_{\gamma} \sim 0.01$ keV, and conversion to continue in the range from $\sim \rm{R}$ to $1.2\rm{R}$. The threshold level $\epsilon_{\rm{p,Th}}$ for $\Delta^{+}$ production in $\rm{p-UV}$(A/B) interaction according to the equation (1) increases rapidly with $\rm{r}$ because of the angle factor $(1-\rm{cos}\theta_{\rm{p}\gamma})^{-1}$. The conversion probability can be parametrized as $\rm{P_{c}} \simeq  \rm{T}^{3}_{0.1 \; \rm{keV}}$ using modified radiative field temperature \citep{lin06}. Therefore, very near to the upper bound of the $\rm{p} \rightarrow \Delta^{+}$ transformation region, the total flux of neutrinos/gamma-rays that is originated from the disintegration of $\Delta^{+}$ resonance state will be 

\begin{equation}  
\Phi_{\nu / \gamma}(\rm{r} \simeq 1.2\rm{R}) =  2 \rm{c f}_{\xi} \rm{A_{pc} f_{d} (1-f_{d}) n_{o} P_{c}} ,
\end{equation}  
  
with $\rm{f}_{\xi}$ is $4/3$ and $2/3$ for $\gamma$-rays and $\nu_{\mu}$s respectively. If now the duty cycle factor $\rm{f_{dc}}$ of the gamma-ray/muon neutrino is taken into account, the phase averaged gamma-ray/$\nu_{\mu}$ flux on the Earth from a magnetar at a distance $\rm D$ is given by 

\begin{equation}
\Phi_{{\nu}_{\mu},\bar{\nu_{\mu}} / \gamma} \simeq  2 \rm{c f}_{\xi} \rm{f}_{\zeta} \eta_{\rm{A}} \rm{f_{dc} f_{s} f_{d} (1-f_{d}) n_{o}}\left(\frac{\rm{R}}{\rm{D}}\right)^{2}\rm{P_{c}}
\end{equation} 

In equation (4), the flavor ratio of neutrino at their production point to a very large distance, say, at a detection level on earth is different due to well-known neutrino oscillations. The effect of neutrino oscillations is represented by the  parameter $\rm{f}_{\zeta}$ ( 1/2 and 1 for muon neutrinos and gamma-rays). The factor $\rm{f_{s}}$ is set equal to 1 for $\nu_{\mu}$ but not yet known correctly for gamma-rays. The factor $2$ comes in equation (3) or (4) due to $\bar{\nu_{\mu}}$ production in the process (A). 

The efficiency of the dynamo process in young magnetars is partly defined by the initial angular speed $\Omega_{\rm i} = 2\pi/\rm{P_{i}}$. The associated analytical calculations suggest that the creation of such enormous magnetic field needs rotational period with $\rm{P_{i}} \sim 1$ ms at birth time \citep{dan92}. Under such a circumstance, stars possess huge rotational energy for a small period before transforming into other class of stars through rapid evolutions. But, in this work we have taken $<\rm{P}>$ as $\sim 350$ ms, so that the magnetar has just passed the neutrino-loud regime during its evolutionary phase \citep{zha03}. 

We now calculate numerical values for $\nu_{\mu}$ and $\gamma$-ray fluxes using the formula in the equation (4) for a typical galactic magnetar with $\rm{D} \sim 2$ kpc, $\rm{P} \sim 350$ ms, $\rm{B}_{15} \sim 1.5$, $\rm{T}_{0.1 \; \rm{keV}} \sim 0.0255$, and $\rm{f_{dc}} \leq 0.10$ in both the cases when $\eta_{\rm{A}}$ is equal to (i) 1 and (ii) $\Omega \rm{R/(4c})$ . The factor $\rm{f_{s}}$ in the equation (4) is set equal to $\sim 1$ in accordance with the explanation pointed out in Sec. II. We have taken star radius equal to $\rm{R} = 10$ km for the present calculation. For the purpose, we choose $\rm{Z} = 1$ and $\rm{f_{d}} = 1/2$ here.

The corresponding $\nu_{\mu}$ and $\gamma$-ray integral fluxes ($\rm E^{2}\phi_{\nu_{\mu}/\gamma}$) calculated out from the equation (4) are $6.03 \times 10^{-10}$  and $48.34 \times 10^{-10}$ in $\rm{GeV cm^{-2}s}^{-1}$ for $\eta_{\rm A} = 1$. These values are $0.0009 \times 10^{-10}$ and $0.007 \times 10^{-10}$ according to the case (ii) in $\rm{GeV cm^{-2}s}^{-1}$. If we compare with IceCube estimated integral PeV neutrino flux, that is, $\sim 2.4\times 10^{-9}$ $\rm{GeV cm^{-2}s}^{-1}$, these predicted values look quite low, particularly in (ii). 

\subsection{The neutrino and gamma-ray energies}

The average percentage of proton energy converted to the pion in the photomeson process is $\sim 20$\%, or in terms of temperature the pion energy would be $\sim 200 \times T^{-1}_{0.1 \; \rm{keV}}$ TeV  \citep{lin05}. Since the pion resides only a very short time in the dense UV radiation zone and hence suffers negligible energy loss through inverse Compton scattering. Subsequently pion decay occurs and $\frac{1}{4}$-th of its energy will be transferred into muon neutrino, the rest will equally be shared by the other three leptons. Average energy of the $\nu_{\mu}$ for a young magnetar with $\rm{T}_{0.1 \; \rm{keV}} \sim 0.0255$ is therefore

\begin{equation}
\epsilon_{\nu_{\mu}} \sim 50 \times \rm{T}^{-1}_{0.1 \; \rm{keV}} \rm{TeV} \sim 1.97 \rm{PeV}.    
\end{equation}

On the other hand, the average energy of $\gamma$-rays is expected to be 
\begin{equation}
\epsilon_{\gamma} \sim 100 \times \rm{T}^{-1}_{0.1 \; \rm{keV}} \rm{TeV} \sim 3.93 \rm{PeV}.   
\end{equation}

But, the resulting $\gamma$-ray energy is expected to be smaller than the model prediction due to QED phenomenon in the region around $1.2\rm R$ stellar distance where magnetic-field intensity reduces to $\rm{B_{cr}} \approx 4.4 \times 10^{13}$ G in magnetars. This energy degradation aspect of gamma-rays is already discussed in Sect. II. It should be however mentioned that in the present calculation the loss due to pion curvature process is considered inefficient \citep{her08}.

\section{PeV neutrino and gamma-ray events in VHE ground based experiments}

\subsection{Detection of PeV neutrino events in IceCube}

Detection of two neutrino events with energies in the range $1- 10$ PeV has been reported by IceCube experiment \citep{aar13}. These high-energy muon neutrinos are usually detected indirectly through the observation of the Cherenkov light produced in ice by charged secondary particles created from neutral-current or charged-current interactions. The visible track length leaving behind the path of a produced high-energy $\mu$ is estimated from the reconstruction of the Cherenkov light detected by optical sensors buried in the Antarctic ice in configuration with the digital optical modules. Events with angular resolution $\leq 1^{\rm o}$, $50\%$ confidence level correspond visible muon tracks out of all events \citep{aara14}. Event selection for astrophysical neutrinos from backgrounds mostly arising from VHE cosmic ray air shower induced muons is implemented through Monte Carlo simulations \citep{abb13, aarb13}. The sensitivity of the IceCube to VHE neutrinos is found above a threshold of $\epsilon_{\nu_{\mu}}^{\rm{\rm{Th.}}} \sim 0.1$ TeV. IceCube with that level of sensitivity has measured neutrino flux as $\rm{E}^{2}\phi_{\nu_{\mu}+{\tilde{\nu}_{\mu}}} \sim 3\times 10^{-8}$ GeV cm$^{-2}$ s$^{-1}$ sr$^{-1}$ corresponding to neutrino energy range $0.2 - 2$ PeV. For the typical magnetar considered in this work, the probability of conversion of ${\nu_{\mu} \rightarrow \mu}$ in ice is expected to be $\rm{p}_{\nu_{\mu} \rightarrow \mu} \simeq 1.3 \times 10^{-6}\left(\frac{\epsilon_{\nu_{\mu}}}{ 1 \; \rm{PeV}} \right)$ \citep{gai95}, and corresponding event rates in IceCube detector would be 

\begin{equation}
\frac{\rm{dN}}{\rm{dAdt}} = {\phi_{\nu_{\mu}}\rm{p}_{\nu_{\mu} \rightarrow \mu}} \leq {2.5 \times 10^{-4}}  {\rm{km}}^{-2}\rm{yr}^{-1}.  
\end{equation}

It is clear that the model dependent calculation for the PeV $\nu_{\mu}$ flux and the event rates obtained in the subsection III D and using equation (7) are much lower for $\eta_{\rm{A}} \neq 1$ compared to IceCube limits with its current sensitivity. It is expected that a next-generation (i.e., {\lq Gen2\rq} with instrumented volume $10$ km$^3$ and $75$ km$^2$ surface array) IceCube detector will be able to cover these lower limits of $\nu_{\mu}$ flux and hence will pinpoint the site of the VHE events in the universe.

\subsection{Detection of PeV gamma-rays in IceCube and HAWC}
 
Cosmic rays (CRs) beyond 1 PeV and perhaps as close as $10^3$ PeV, are believed to be of galactic origin. Therefore, detection of PeV gamma-rays provides an opportunity to know the possible origin and acceleration mechanisms of high-energy CRs. For observing TeV gamma-rays, the Imaging Air Cherenkov Telescopes (IACTs) and air fluorescence detectors have been used in many experiments \citep{abra10}. High-energy gamma-rays might also be detected by ground based air-shower arrays equipped with muon detectors \citep{gup05}. The PeV energy scale is inaccessible to even present generation IACTs viz. H.E.S.S. \citep{aha06}, MAGIC \citep{ali08} and VERITAS \citep{cel08}. Under the present perspective of the work, the simultaneous observation of muon neutrinos and gamma-rays at PeV energies has to be ensured in IceCube experiment \citep{gup13}. A new approach combining muon-poor air shower data from the IceCube's surface component i.e., IceTop with the in-ice array data of IceCube may identify the PeV gamma-ray events \citep{aarc13}. IceCube's 5-year sensitivity to point sources indicates that if the spectral index remains unchanged across the energy range from $10$ TeV to $\sim 1$ PeV, the gamma-ray flux would have been measured up to a lower limit of the order of $\sim 10^{-10}$ GeV cm$^{-2}$ s$^{-1}$ at 1 PeV from regions at $5-8$ kpc distance scale. This clearly suggests that the present sensitivity of IceCube to PeV gamma-rays from magnetars with $\eta_{\rm A} \neq 1$ is still not within the reach by the system. Due to very low flux obtained from the model calculation, IceCube therefore does not hint about any hypothetical detectable magnetar candidate even as an unidentified source in near future in the existing gamma-ray data sets.  

The upcoming planned Cherenkov Telescope Array (CTA) \citep{ach13} with a factor of 5-10 times higher in sensitivity compared to current generation IACT instruments is expected to span from a few tens of GeV to around 100 TeV. Hence, observation of gamma-rays in the PeV energy scale by the CTA does not look bright. 

In recent times, the High Altitude Water Cherenkov (HAWC) gamma-ray observatory has already shown its potentiality to detect high energy gamma-rays up to 100 TeV \citep{abe12}. If the HAWC keeps its sensitivity to VHE gamma-rays unchanged in the region 10 TeV - 1 PeV following a constant spectral index, the possible lower bound of gamma-ray flux results into $\sim 4 \times 10^{-11}$ GeV cm$^{-2}$ s$^{-1}$ for sources within 10 kpc. In comparison with the PeV gamma-ray flux obtained from a magnetar at $\sim 2$ kpc distance with $\eta_{\rm{A}} \neq 1$ ($\sim 0.007 \times 10^{-10}$ GeV cm$^{-2}$ s$^{-1}$), the HAWC results under some principle conditions casting some sign of improvement in the VHE gamma-ray astronomy.      

The daughter particles from $\rm{p}\gamma$ interactions are neutrinos, gamma-rays, degraded protons and neutrons. Among them protons may remain trapped in star's magnetic field and neutrons will decay and result into CRs. Hence these components cannot reach the earth uninterrupted due to various interactions and deflections in the interstellar medium, and thereby obscuring directional information completely. In each such $\rm{p}\gamma$ interaction, VHE protons lose about $20\%$ of their initial energy from which $\nu_{\mu}$ and $\gamma$-ray gain $\sim 5\%$ of the proton energy each. This suggests a possibility of simultaneous detections of PeV neutrinos and gamma-rays from magnetars on earth if the gamma-rays escape the emission region without making pairs (more likely since mean free path for $\rm{e}^{+} - \rm{e}^{-}$ pair production is $\geq 10$ kpc \citep{pro96}). High sensitive IceCube system and HAWC experiments in future may be useful to detect these PeV gamma-rays \citep{aarc13,abe13}. These PeV gamma-rays are unlikely to reach at detection site from extragalactic sources because of their pair-production losses with cosmic background radiations. Hence, results from HAWC on the VHE gamma-ray component and the simultaneous investigation for PeV neutrinos and gamma-rays by the upcoming Icecube's Gen2 with higher sensitivity will resolve the enigma on the origin of PeV gamma-rays in coming years. 

A simultaneous generation of PeV neutrinos and gamma-rays in the magnetar model follows some important astrophysical processes which are given in (A). These processes altogether refer to an important viewpoint that PeV neutrinos and gamma-rays from magnetars are hadronic in fundamental origin. There is a recent prolific observational evidence on VHE gamma-ray emission from the Crab pulsar reaching up to 1.5 TeV energy. Such TeV gamma-ray observations strongly favor inverse Compton (IC) scattering off low-energy photons in the wind acceleration region very close to the light cylinder \citep{ans16}. This clearly refers to another important viewpoint in which these gamma-rays may well be considered to be leptonic in fundamental origin. However, these TeV gamma-rays do not accompany any neutrinos at their birth from the production site. Absence of neutrinos and the upper limit of energy (up to 1.5 TeV only) bring important constraints on the functioning of their model at larger distances (where the fields will presumably be weaker) in the present scenario. 

\section{Conclusions}

If protons reach $10 - 100$ PeV energy scale in a magnetar then their interactions with modified UV-A/B photons may generate PeV neutrino events with energies between $1 - 10$ PeV as observed by the IceCube experiment. The predicted event rates of PeV neutrinos by the model suggest no possible indication of any statistically significant excess from the direction of any local magnetar to be observed by IceCube in near future and is thus as per IceCube expectations \citep{aard14}.

Since the process of photon splitting in the PeV region is not studied yet in QED, and if PeV gamma-rays might have generated in the magnetosphere where $\rm{B > B_{cr}}$, then one cannot rule out the modification of the PeV gamma-ray spectra detected at earth from magnetars.

If gamma-rays are detected simultaneously with neutrinos at PeV energies then the origin of PeV astrophysical neutrinos would probably be resolved.   

\section*{Acknowledgements}
\noindent The authors would like to thank Prof. Dmitry Rumuyantsev and Prof. Mikhail Chistyakov of Yaroslavl State (P.G. Demidov) University, Russia for helpful discussions on some particular features of the photon splitting process in \citep{chi12}. RKD thanks Dr A. Bhadra, HECRRC, NBU for some fruitful discussion on the magnetar model. RKD acknowledges the financial support from SERB, Department of Science and Technology (Govt. of India) under the Grant no. EMR/2015/001390.

\end{document}